\lineskip=3pt minus 2pt
\lineskiplimit=3pt
\magnification=1200

\def\hf{\hfil\break}
\def\pmb#1{\setbox0=\hbox{$#1$}%
 \kern-.025em\copy0\kern-\wd0
 \kern.05em\copy0\kern-\wd0
 \kern-.025em\raise.0433em\box0 }
\centerline{\bf Evidence for an Intense Neutrino Flux during $r$-Process
Nucleosynthesis?}
\vskip 24pt
\centerline{{\sl Gail C. McLaughlin} and {\sl George M.
Fuller}}
\vskip 0.2in
\centerline{Department of Physics}
\centerline{University of California, San Diego}
\centerline{La Jolla, CA 92093-0319}
\vskip 0.6in
\centerline{\sl ABSTRACT}
\vskip 0.2in
We investigate the possibility that neutrino capture on heavy nuclei competes
with beta decay in the environment where the $r$-Process elements are
synthesized.  We find that such neutrino capture is not excluded by existing
abundance determinations.  We show that inclusion of significant neutrino
capture on the (neutron number) N=82 waiting point nuclei can allow the inferred
abundances of these species to provide a good fit to steady weak (beta decay plus
neutrino capture) flow equilibrium.  In fact, for particular choices of neutrino
flux conditions, this fit is improved over the case where nuclei change their
charge by beta decay alone.  However, this improved fit can be realized only if
neutrino capture plays a negligible role in nuclear decay back toward 
stability.  We discuss the implications of these considerations for
current proposed sites and models for $r$-Process nucleosynthesis.
\vskip 0.2in
{\it Subject Headings}: elementary particles - 
nuclear reactions, nucleosynthesis,
abundances - stars: supernovae
\vfil
\eject
\vskip 0.2in
\noindent
\centerline{\bf{1. Introduction}}
\vskip 0.2in
Many elements heavier than the iron peak, and about half of nuclei heavier than mass $A >
100$, are believed to be synthesized in the $r$-Process (or rapid neutron
capture process; Burbridge et al. 1957; Cameron et al. 1957; For a recent
review see Meyer 1994).
 The $r$-Process is thought to 
take place in conditions of (n,$\gamma$), ($\gamma$,n) equilibrium. (cf. Kratz et
al. 1988, Kratz et al. 1993).  Measurements of abundances of r-process nuclei 
show
that peaks exist around the closed neutron shells, occurring at neutron
numbers N=50, N=82, and
N=126.  In particular, the relative abundances of the $r$-Process
progenitor elements
can be determined to accuracies of order $\sim 20\%$ in the vicinity of
the N=82 (nuclear mass number 
${\rm A}\approx 130$) abundance peak (Kratz et al. 1988).  
This determination involves 
first subtracting the predicted s-process contribution from the measured solar
abundance for each stable nuclide. 
The $r$-Process progenitor abundances then are inferred from beta decay
lifetimes and the estimated 
probabilities for beta-delayed neutron emission along the decay path back to
stability.  

In (n,$\gamma$), ($\gamma$,n) equilibrium, the abundance distribution along an
isotopic chain often can be strongly peaked at one or a few nuclear species
around the closed neutron shells.  We define $Y(Z)$ to be the abundance of the
nuclide at the abundance peak for an isotopic chain Z.  In the limit where
$Y(Z)$ is approximately equal to the total abundance along isotopic chain $Z$, 
the dominant source of leakage from isotopic chain Z to chain $Z+1$ will occur
through the beta decay of this nucleus at the abundance peak.  
We define the beta decay
rate of this nuclide to be $\lambda_\beta(Z)$.  Usually, such a \lq\lq waiting
point\rq\rq\ nucleus will have a closed neutron shell.  
Conditions of local steady
nuclear flow (time independent abundances) would correspond to a constant value
of the product $Y(Z) \lambda_\beta(Z)$ over some range of nuclear charges.  This
is sometimes termed steady beta flow equilibrium (cf. Cameron, Cowan \& Truran
1983).  We can define the beta flow ratio $R_\beta(Z,Z^\prime)$ for isotopic
chains $Z$ and $Z^\prime$ to be,
$$ R_\beta(Z,Z^\prime) \equiv {Y(Z) \lambda_\beta(Z) \over Y(Z^\prime) 
\lambda_\beta (Z^\prime)}. \eqno(1)$$
If $Z$ and $Z^\prime$ are within the range of nuclear charges where a local
steady beta flow obtains, then we would expect $  R_\beta(Z,Z^\prime) = 1$. 
Obviously, these considerations for defining steady beta flow would have to be
modified if there are two or more isotopes which are significantly populated in
a given isotopic chain.

Conditions of steady beta flow and $(n,\gamma), (\gamma,n)$ equilibrium could
occur in many of the diverse environments which have been suggested as possible 
sites of origin 
for $r$-Process nucleosynthesis (for reviews see Mathews \& Cowan 1990; Cowan,
Thielemann \& Truran 1991; and Meyer 1994; and references therein).  However, two
proposed $r$-Process sites could occur in intense neutrino fluxes: 
decompression of cold neutron matter from neutron star mergers (cf. Meyer 1989;
Lattimer et al. 1977); and neutrino heated supernova ejecta (Woosley \&
Hoffman 1992; Meyer et al. 1992; Woosley et al. 1994; Janka, Takahashi \& Witti
1994).  The processes of neutrino and antineutrino capture on heavy nuclei:
$${\nu}_e + {\rm A(Z,N)} \rightarrow  {\rm A(Z+1,N-1)} +  
e^{-};\eqno(2a)$$
$${\bar{\nu}}_e + {\rm A(Z,N)} \rightarrow  {\rm A(Z-1,N+1)} +  
e^{+},\eqno(2b)$$
produce the same nuclear charge changing effect as beta decay (positron capture)
and positron decay (electron capture), respectively, though the nuclear dynamics
and energetics of these processes can be very different (cf. Fuller \& Meyer,
1995 hereafter FM95; McLaughlin \& Fuller, 1995 hereafter MF).  
For the neutron-rich nuclei which are of
interest in the $r$-Process and for the neutrino and antineutrino energy spectra
expected from hot nuclear matter, only neutrino capture (equation 2a) could play
an important role (FM95; MF).  
If the rates of
neutrino capture on the waiting point nuclei in the $r$-Process compete favorably
with the corresponding beta decay rates, then we could expect significant
alterations of our picture of steady flow equilibrium (FM95;
Nadyozhin \& Panov 1993).

Neutrino capture on heavy nuclei could be important if the neutrino flux is large
and/or if the neutrinos have high energies.  The neutrino
capture rate, $\lambda_\nu$, depends both on nuclear physics and on the 
properties of the
neutrino flux.  For example, 
assuming that the neutrino flux originates from a spherical hot proto-neutron
star, the following rough dependence holds
 (FM95; MF):
$$ \lambda_\nu \propto L_\nu T_\nu r_7^{-2} |M|^2 P. \eqno(3)$$
Here, $L_\nu$ is the energy 
luminosity of the neutrinos, $T_\nu$ is the temperature
of the electron neutrino distribution function (assuming the neutrino chemical
potential is $\mu_\nu \approx 0$),
 and $r_7$ is the distance from the center of the neutron star in units of
$10^7 {\rm cm}$.  The characteristic 
nuclear matrix element is $|M|^2$, and $P$ is a characteristic phase 
space factor.  Given a sufficiently energetic neutrino distribution function, we
expect neutrino capture to be dominated by
transitions to the Fermi and Gamow-Teller Resonance states (cf. Figure 1 of FM95).
  These weak strength distributions peak at excitation energies
which are at least the coulomb energy difference between the parent and
daughter nuclei plus the parent-daughter mass difference.  We caution that the
dependence
of $\lambda_\nu$ on the various ingredient quantities in equation (3) is highly
dependent on the particular geometry of our example (neutrino-heated supernova
ejecta from a hot proto-neutron star).  Alternatively, if we were to consider
outflow from merging neutron stars, then we could obtain similar values of
$\lambda_\nu$ and $r$-Process effects of neutrino capture, though the
dependence on the geometry and neutrino distribution function parameters could be
significantly different than in the supernova case.

\vskip 0.2in
\centerline{\bf{2. Steady Beta Flow vs. Steady Weak Flow}}
\vskip 0.2in
We have calculated the ratios $R_\beta$ for elements in the N = 82 peak. 
The beta decay rates we used (Kratz et al. 1988; Tuli 1990) 
are shown in Table 1.  In general, the heavier
elements 
tend to have slower beta decay rates since these also have larger Z, and thus 
are closer to stability.  The abundances used (Kratz et al. 1988) are shown
in Table 1. The
resulting values of $R_\beta$ are shown in Table 2.  The $R_\beta$ ratios
 have been calculated 
with the heavier (more stable) element always in the numerator, and the lighter
element always in the denominator.  These values fall systematically below
unity.  This systematic trend has not been pointed out before.
However, the $R_\beta$ values in Table 2 usually are taken as evidence
for steady beta flow in the $r$-Process, since the characteristic abundance
errors are believed to be of order $15-20\%$ and the $R_\beta$ values differ
from unity by roughly that amount.  Unless there are systematic, non-Gaussian
normal errors in the abundance determinations, it is difficult to understand why
these $R_\beta$ ratios are systematically low.  Although most of the
elements in the abundance peak are concentrated in the N=82 isotopes, a small 
component in
another isotope will cause a deviation of the ratios $R_\beta$ from unity.
Below we will argue that $\nu$-capture could explain this result, however, 
Thielemann (1996) has suggested
that a superposition of several distinct neutron exposure conditions could
account for the systematic trend evident in Table 2.  There 
remains the
question of whether the particular temperature, density, and $Y_e$
(electron fraction) tracks proposed in, for example, the Woosley et al. (1994)
or the Qian \& Woosley (1996) ejection trajectories provide the requisite
superposition of conditions to explain the trend.  Note that we have employed ground state beta
decay rates in Tables 1 and 2.  If the $r$-Process is occurring in high temperature environments,
then there may be an 
acceleration of the beta decay rates resulting from thermal population of excited
nuclear parent states.  

However, we note that a modicum of neutrino exposure can lead
to an interesting interpretation of the trends in Table 2.  To take account of
the possibility that neutrino capture, as well as beta decay, contribute to the
rate of weak leakage from one isotopic chain to the next, we define, 
$$ R_{\beta + \nu} (Z,Z^\prime) \equiv {Y(Z) (\lambda_\beta(Z) + \lambda_\nu(Z))
 \over Y(Z^\prime) (\lambda_\beta (Z^\prime) + \lambda_\nu (Z^\prime))}, 
\eqno(4)$$
where $\lambda_\nu(Z)$ and $\lambda_\nu (Z^\prime)$ are the neutrino capture
rates on the waiting point nuclei of charge $Z$ and $Z^\prime$, respectively,
while the other notation is as in equation 1.  Clearly $ R_{\beta + \nu} = 1$ if
steady flow equilibrium prevails over the region of the abundance peak which
includes $Z$ and $Z^\prime$.  We will term this condition steady {\it weak} flow
equilibrium.

The evaluation of $R_{\beta + \nu}(Z,Z^\prime)$ is more difficult than the 
calculation of $R_{\beta}(Z,Z^\prime)$.
The neutrino
capture rate will vary according to the properties of the neutrino flux and
distance from the neutron star or other neutrino source.  Also, the rate will 
vary between nuclei due to
differences in characteristic matrix elements
 and the excitation energy of the Fermi and 
Gamow-Teller resonances.  Our calculations of the neutrino capture rates 
for the $N=82$ waiting point nuclei have been tabulated in Table 1.  
These computations
 of the neutrino capture cross sections conform to the method
of FM95 and MF, except we here
include a full numerical calculation of the energy dependent phase space factors
and Coulomb wave correction factors (McLaughlin \& Fuller 1996).  For these specific rates in Table
1, a neutrino luminosity of $L_\nu =  10^{51} {\rm ergs \, s^{-1}}$ and a $\nu_e$
neutrino sphere 
temperature of $T_\nu = 3.1 \, {\rm MeV}$ is assumed.  These latter 
conditions are
 consistent
with those obtained at late time $(t_{pb} \approx 7s)$,  where most of the
N=82 nuclei are made (Woosley et al. 1994).  
Using these conditions, and a distance of $r_7 = 0.8$, we find
that the ratio $R_{\beta + \nu} \approx 1$ in all cases.  
 The $R_{\beta + \nu}$
ratios for  $r_7 = 0.8$ are shown in
Table 3.  These results demonstrate that significant neutrino capture could be 
tolerated in some models of 
$r$-Process nucleosynthesis.  Note that the neutrino flux parameters and the 
distance to the neutron star may be varied as long as the value 
$L_\nu T_\nu r_7^{-2}$ remains the same, at least in the context of the model
of $r$-Process nucleosynthesis 
from supernova outflow.  If the radius is increased by a factor 
of 2, the luminosity or
temperature must decrease by a factor of 4.  Other possible alterations to the
capture rates are discussed below.

Additional variance in neutrino capture rates between nuclei could be found
through differences in the
placement and width of the Gamow-Teller strength.  The 
excitation energy distribution 
of this strength is uncertain (FM95, MF).  
The Gamow-Teller strength distribution width  in the
 ${\rm N} = 82$ nuclei may be varied to effect an 
increase in the neutrino capture rate in these species 
by a factor of about $\sim 2$. 
The neutrino capture rate calculations presented
 in Table 1 are based on the assumption that the Gamow Teller resonance
was located at the same excitation energy as the Fermi strength.  If the
Gamow-Teller strength centroid 
is pushed upward by about 4 MeV in excitation energy,
 then the neutrino capture
rate could decrease by as much as a factor of $\sim 2$.  
Such effects could cause significant rate
variations between nuclei.  In the case of the nuclei examined in Tables
2 and 3, all $R_{\beta + \nu}$ ratios are close to unity when calculated with
 the given neutrino flux parameters. 
If the capture rate for one nucleus were to change (e.g. from a change in the
width or centroid of the Gamow-Teller strength), then all the ratios $R_{\beta + \nu}$
would be driven further from unity when an attempt is made to find a new
concordant $r_7$ value.

\vskip 0.2in
\centerline{\bf{3. Discussion}}
\vskip 0.2in
First, we must note that in
the particular environment of the post-core-bounce Type II supernova, an
r-Process at $r_7 \sim 1$ is in conflict with the models based on the
high entropy bubble in the
Wilson and Mayle supernova calculations.  At the 
high entropy per baryon $s/k \approx 400$ obtained in these models, the
$r$-Process goes on at, or above, about 600 km - 1000 km 
(Woosley et al. 1994; Meyer et al. 1992).
However, if the entropy at this time were lower, $s/k \approx 100$, then heavy
element nucleosynthesis ($(n,\gamma), (\gamma,n)$ equilibrium) 
could conceivably occur much closer to the neutron star, with  
$r_7 \sim 1$ not out of the question
(see for example the calculations of Qian \& Woosley 1996 where it is argued
that the entropy must be less than $s/k \approx 200$).  
Production of r-process elements in this scenario requires
further investigation into the velocity field and electron fraction of the
outflowing material.  Other numerical supernova models (e. g. the models on which
the Takahashi, Witti \& Janka 1994 calculation are based) seem to get such low
entropies.  Burrows, Hayes \& Fryxell (1995) also seem
 to obtain low entropy, though it might be
argued that these calculations do not probe the late epochs where the
$r$-Process might occur.  In any case, it is not obvious that any of the existing $r$-Process
models based on neutrino-heated supernova ejecta could give $(n,\gamma),(\gamma,n)$ equilibrium
freeze-out for the ${\rm N} = 82$ nuclei at $r_7 \sim 1$. This implies that either: (1) the trend
in Table 2 does not arise from neutrino capture; (2) models of the $r$-Process from supernova
ejecta need to be altered; or (3) the $r$-Process originates in some other site.

In the Woosley et al. (1994) calculations, mass elements (trajectories in their terminology) which
leave the neutron star at progressively later times tend to be responsible for the production of
progressively heavier $r$-Process nuclear mass ranges.  It is clear that neutrino capture can
dominate over beta decay throughout the stage of the $r$-Process leading up to the establishment of
$(n,\gamma),(\gamma,n)$ equilibrium.  Indeed, this can accelerate the $r$-Process, (Nadyozhin \&
Panov 1993), and perhaps explain how the \lq\lq scale height\rq\rq\ of models could be reduced.
However, neutrino capture must be comparable or subdominant to the beta decay rates
of the waiting point species when freezeout from $(n,\gamma),(\gamma,n)$-equilibrium occurs.

It would be desirable to examine steady weak flow in
the N=50 peak nuclei in a manner similar to our treatment of the 
N=82 nuclei.  
Unfortunately, due to difficulties in extracting the $r$-Process component of
the measured abundances for these species, 
the inferred progenitor abundances (e.g. $^{80}{\rm Zn}$), are 
more uncertain than those inferred for the N=82 abundance peak 
(Kratz et al. 1988).  This makes analysis of the weak flow
 in this region 
difficult.  Fuller \& Meyer (1995, 1996) considered the ratio of $^{80}{\rm Zn}$ 
to $^{79}{\rm Cu}$. They found
that if steady beta flow is assumed ($R_\beta = 1$ to within 20\%), then a
significant rate of neutrino capture on heavy nuclei is not tolerable.  In fact,
they employed their calculations to place a limit of  
$r_7 > 4$ on the location of $(n,\gamma),(\gamma,n)$- equilibrium freeze-out
for the ${\rm N}=50$ nuclei.
This limiting value of radius depends quite sensitively on the value 
of $R_\beta$. 
Therefore, it would be desirable to know a more precise
 value of the progenitor abundances in order to make a
determination of the prospects for significant neutrino capture contributions to
steady weak flow in
this case.  Furthermore, in models of neutrino-heated supernova ejecta, the N=50
nuclei are primarily made much earlier than are the 
${\rm N}=82$ species.  The
neutrino-heating history, outflow velocity, and other parameters could be quite
different at this earlier epoch.  Therefore, it is conceivable that the
Fuller \& Meyer (1995, 1996) limit on the location of the 
${\rm N}=50$ freezeout could be
consistent with significant neutrino capture in the
 ${\rm N}=82$ species, at least within the context of
the $r$-Process originating in supernova ejecta.

If neutrino capture can affect steady weak flow equilibrium, then would it not
also play a role during the decay back toward stability?  Neutrino
capture-induced neutron emission would be greatly enhanced over beta-delayed
neutron emission, since neutrino capture could access the considerable
Gamow-Teller and Fermi strength which resides at daughter nucleus excitation
energies above the neutron separation energy for the $r$-Process progenitor
species (FM95).  Indeed, populating the region of excitation
energy near the isobaric analog state or Gamow-Teller peak could result in
multiple neutron emission.  Clearly, if such processes operate during decay back
toward stability in the $r$-Process, then the inference of the progenitor
abundances from the measured stable species abundances could be quite
significantly different from the results of Kratz et al. (1988, 1993).  In this 
event, our discussion of neutrino capture contributions to weak steady flow
based on the Kratz et al. (1988, 1993) progenitor abundances would be specious,
and the remarkable results of Table 3 merely an accident.  For this scenario,
perhaps the Thielemann (1996) suggestion for explaining the trends in Table 2
will prove to be correct, in which case neutrino capture will serve to drive the
$R_{\beta + \nu}$ ratios further from unity.  Such an effect could allow for
very useful and stringent constraints on the location of the $(n,\gamma),
(\gamma,n)$-equilibrium freezeout in various models for the $r$-Process.

However, the rapid outflow of material inherent in some models of the
$r$-Process suggests an alternative resolution to the inconsistency of
significant neutrino capture in both weak steady flow and during decay back
toward stability.  As a working example, let us consider models of the
$r$-Process inspired by the rapidly outflowing, neutrino-driven wind in
the post-core-bounce supernova environment.  Here, we would expect neutrino capture
capture and decay back toward stability to be occurring in material which is
moving away from the source of the neutrino flux.  If indeed neutrino
capture is as significant in steady weak flow as Table 3 suggests, 
then we know that
the neutrino flux must be high enough to give the neutrino capture rates in
Table 1 for the N=82 species when this material freezes out of
$(n,\gamma),(\gamma,n)$ equilibrium.  The condition 
$R_{\beta + \nu} \approx 1$ in
Table 3 suggests that the $N=82$ nuclei experience this freeze-out at $r_7 \approx
0.8$.  If neutrino capture is to play a negligible role in decay back toward
stability, then conservatively the mean neutrino capture rates should be less
than about one tenth of the beta decay rates.  So, for example, for $^{127}{\rm
Rh}$, $\lambda_\nu \approx 7.4 \, {\rm s}^{-1}$ at $r_7 = 0.8$, while $\lambda_\beta
\approx 10.35 \, {\rm s}^{-1}$.  A characteristic outflow velocity of $v \approx
10^8 {\rm cm \, s}^{-1}$ will take the material at the $(n,\gamma),(\gamma,n)$
freeze-out point at $r_7 \approx 0.8$ to $r_7 \approx 2$ in one neutrino capture
time.  At $r_7 \approx 2$, we will have $\lambda_\nu \ll \lambda_\beta$.  An
outflow velocity $v \geq 10^8 {\rm cm \, s}^{-1}$ is just what is expected in
recent models.  Therefore, the rapid outflow of the material could
allow significant neutrino capture influence on steady flow equilibrium, yet
minimize neutrino-induced processing during decay back.

Neutrino capture conceivably could populate highly excited states in the daughter nucleus which lie
above the fission barrier.  Such neutrino capture-induced fission will be most likely for heavier
nuclei, including the ${\rm A} \approx 195$ peak nuclei, and especially the actinides (Fuller,
McLaughlin, \& Meyer 1996).

Could neutrino capture-induced fissioning of heavy nuclei halt the $r$-Process
flow before the actinides could be synthesized?  Again, rapid outflow could
allow lower mass nuclei to experience significant neutrino flux exposure, while
the heavier species which are synthesized later experience insignificant
neutrino processing.  Although this is plausible, it remains to be seen whether
detailed models of the $r$-Process in neutrino-heated outflow can avoid problems
with neutrino capture-induced fission or neutrino processing on the decay back
to stability.

\centerline{\bf{Acknowledgements}}

We wish to thank B. S. Meyer, F.K. Thielemann, Y. Z. Qian and S. E. Woosley
 for useful discussions.  This work
was supported by NSF Grant PHY-9503384 and a NASA theory grant at UCSD.
\vfil
\eject

\noindent
\centerline{\bf Table 1: N=82 Nuclei}    
\vskip 0.2in
\noindent
$$\vbox{\offinterlineskip
\hrule
\halign{&#&\strut\quad\hfil#\quad\cr
& &&$^{130}{\rm Cd}$  
 \hfil&&$^{129}{\rm Ag}$ 
\hfil&&$^{128}{\rm Pd}$ \hfil&&
$^{127}{\rm Rh}$\hfil&\cr
\noalign{\hrule}
&$\lambda_\beta$&&$3.47^a$&&$4.88^a$&&$6.03^b$&&$10.35^b$&\cr
&$\lambda_{\nu}$&&$4.01$&&$4.25$&&$4.49$&&$4.75$&\cr
&$(\lambda_\nu / \lambda_\beta) r_7^2$&&$1.15$&&$0.87$&&$0.75$&&$0.46$&\cr
&$Y$&&$2.28^c$&&$2.05^c$&&$1.76^c$&&$1.24^c$&\cr}
\hrule}$$
\vskip 0.2in
\noindent
$^a$ beta decay rates known experimentally from Tuli (1990)

\noindent
$^b$ beta decay rates calculated by Kratz et al. (1988)

\noindent
$^c$ progenitor abundances from Kratz et al. (1988)  
\vfil\eject

\noindent
\centerline{\bf Table 2: $R_{\beta}$ Values$^a$}    
\vskip 0.2in
\noindent
$$\vbox{\offinterlineskip
\hrule
\halign{&#&\strut\quad\hfil#\quad\cr
&  
 \hfil&&$^{129}{\rm Ag}$ 
\hfil&&$^{128}{\rm Pd}$ \hfil&&
$^{127}{\rm Rh}$\hfil&\cr
\noalign{\hrule}
&$^{130}{\rm Cd}$&&$0.79$&&$0.74$&&$0.62$&\cr
&$^{129}{\rm Ag}$&& &&0.92&&$0.77$&\cr
&$^{128}{\rm Pd}$&& && &&$0.82$&&\cr}
\hrule}$$
\vskip 0.2in
\noindent
$^a$ note that the ratios always contain the heavier nucleus' abundance and beta decay rate in the
numerator and the lighter nucleus' abundance and beta decay rate in the denominator.
\vskip 1in
\noindent
\centerline{\bf Table 3: $R_{\beta + \nu}$ Values$^b$}    
\vskip 0.2in
\noindent
$$\vbox{\offinterlineskip
\hrule
\halign{&#&\strut\quad\hfil#\quad\cr
&  
 \hfil&&$^{129}{\rm Ag}$ 
\hfil&&$^{128}{\rm Pd}$ \hfil&&
$^{127}{\rm Rh}$\hfil&\cr
\noalign{\hrule}
&$^{130}{\rm Cd}$&&$0.93$&&$0.95$&&$1.0$&\cr
&$^{129}{\rm Ag}$&& &&1.0&&$1.1$&\cr
&$^{128}{\rm Pd}$&& && &&$1.0$&&\cr}
\hrule}$$
\vskip 0.2in
\noindent
$^b$ neutrino flux evaluated at: $T_\nu = 3.1 {\rm MeV}$, $L_\nu = 10^{51} {\rm ergs \, s^{-1}}$,
and $r_7 = 0.8$
 \vfil\eject

\centerline{\bf{References}}
\vskip 0.2in
\item{}
Burbridge, E. M., Burbridge, G. R., Fowler, W. A., and Hoyle, F. 1957, 
\hf \indent Rev. Mod. Phys., 29, 694

\item{}
Burrows, A., Hayes, J., \& Fryxell, B.A., 1995, ApJ, 450, 830

\item{}
Cameron, A. G. W. 1957, Chalk River Report CRL-41, Atomic Energy Can. Ltd.

\item{}
Cameron, A. G. W., Cowan, J.J., \& Truran, J.W., 1983, Ap\&SS, 91, 235

\item{}
Cowan, J.J., Thielemann, F.-K., \& Truran, J.W., 1991, Phys. Rep. 208, 267

\item{}
Fuller, G. M., McLaughlin, G. C. \& Meyer, B. S., 1996, in preparation

\item{}
Fuller, G. M., \& Meyer, B. S. 1995, ApJ, 453, 792 (FM95)
\par

\item{}
Fuller, G. M., \& Meyer, B. S. 1996, ApJ, in press (erratum to FM95)

\item{}
Kratz, K.-L., Thielemann, F.-K., Hillebrandt, W., M\"oller, P.,
H\"arms, V., W\"ohr, A., \hf
\indent \& Truran, J. W. 1988, J. Phys. G 24, S331
\par

\item{}
Kratz, K.-L., Bitouzet, J.-P., Thielemann, F.-K., Pfeiffer, B. 1993,
ApJ, 403, 216
\par

\item{}
Lattimer, J. M., Mackie, F., Ravenhall, D. G., Schramm, D. N., 1977 ApJ 213, 225

\item{} 
Mathews, G.J., \& Cowan, J. J., 1990, Nature 345, 491
\par

\item{}
McLaughlin, G. C., and  Fuller, G. M. 1995 ApJ, 455, 202 (MF)
\par

\item{}
McLaughlin, G. C. and Fuller, G. M. 1996, in preparation
\par

\item{} Meyer, B. S., 1989, ApJ, 343, 254
\par

\item{}
Meyer, B. S., 1994, Ann. Rev. Astron. Astrophys., 32, 153
\par

\item{}
Meyer, B. S., Mathews, G. J., Howard, W. M., Woosley, S. E.,
\& Hoffman, R. 1992,\hf\indent ApJ, 399, 656
\par

\item{}
Nayozhin, D. K. \& Panov, I. V., 1993, in Proc. Int. Symp. on Weak and
\hf \indent
Electromagnetic Interactions in Nuclei (WEIN-92) ed. Ts. D. Vylov (Singapore:
\hf \indent
World Scientific), 479

\item{}
Takahashi, K., Witti, J., and Janka, H.T., 1994, Astron and Astro. 286, 857
\par

\item{}
Thielemann, F.-K., private communication, 1995
\par

\item{}
Tuli, J. 1990, Nuclear Wallet Cards (Brookhaven: Brookhaven National Laboratory)

\item{}
Qian, Y.-Z., and Woosley, S. E. 1996, ApJ, in press
\par

\item{}
Woosley S. E. \& Hoffman, R. D., 1992, ApJ, 395, 202
\par

\item{}
Woosley, S. E., Wilson, J. R., Mathews, G. J., Hoffman, R. D.,
\& Meyer, B. S. 1994,\hf\indent
ApJ, 433, 229
\par

\vfil
\eject
\end